\documentclass[twocolumn]{aa}
\usepackage{graphicx}
\usepackage{amsmath}

\newcommand{\be}{\begin{equation}}
\newcommand{\ee}{\end{equation}}
\newcommand{\ba}{\begin{eqnarray}}
\newcommand{\ea}{\end{eqnarray}}

\begin{document}

\authorrunning{Geppert, K\"uker \& Page}
\titlerunning{Temperature distribution in magnetized neutron 
              star crusts}

\title{Temperature distribution in magnetized neutron 
              star crusts}
\author{U. Geppert\inst{1} \and M. K\"uker\inst{1} \and Dany Page\inst{2}}
\institute{Astrophysikalisches Institut Potsdam \\
        An der Sternwarte 16 \\
        D-14482 Potsdam, Germany \\
	\and
        Instituto de Astronom\'{\i}a, UNAM, \\
        04510 Mexico D.F., Mexico}
\offprints{U. Geppert, \email{urme@aip.de}}	
\date{}

\abstract{
We investigate the influence of different magnetic field configurations on 
the temperature distribution in neutron star crusts. 
We consider axisymmetric dipolar fields which are either restricted to
the stellar crust, ``crustal fields'', or allowed to penetrate the core,
``core fields''.
By integrating the two-dimensional heat transport equation in the crust,
taking into account the classical (Larmor) anisotropy of the heat conductivity,
we obtain the crustal temperature distribution, assuming an isothermal core.
Including quantum magnetic field effects in the envelope as a boundary condition,
we deduce the corresponding surface temperature distributions.
We find that core fields result in practically isothermal crusts unless the
surface field strength is well above $10^{15}$ G while for crustal fields
with surface strength above a few times $10^{12}$ G significant deviations 
from isothermality occur at core temperatures inferior or equal to $10^8$ K.
At the stellar surface, the cold equatorial region produced by the quantum suppression 
of heat transport perpendicular to the field in the envelope, present for both core and 
crustal fields, is significantly extended by the classical suppression at higher 
densities in the case of crustal fields.
This can result, for crustal fields,
in two small warm polar regions which will have observational consequences:
the neutron star has a small effective thermally emitting area and the
X-ray pulse profiles are expected to have a distinctively different shape
compared to the case of a neutron star with a core field.
These features, when compared with X-ray data on thermal emission of young
cooling neutron stars, will open a way to provide observational evidence in
favor, or against, the two radically different configurations of
crustal or core magnetic fields.

}

\maketitle

\section{Introduction}

The presence of strong magnetic fields in neutron stars is one of 
their distinctive characteristics. 
In typical neutron stars the observed and/or inferred surface 
fields are of the order of $10^{12...13}$ G ; 
for magnetars they even reach $10^{14...15}$ G. 
The strength, structure and time evolution of the magnetic field is
intimately related to its origin, which is still an open problem.
Basically, two qualitatively different types of field structures are 
conceivable, one having an initial field penetrating the whole star 
while the other is characterized by having the field and its 
supporting currents restricted to the stellar crust
(see, e.g., Chanmugam \cite{C92}).
To date, there is still no compelling observational evidence in favor
of or against either of these two hypotheses but recently Link \cite{L03}
argued that long period pulsar precession, as observed in 
PSR B1828-11 (Stairs et al. \cite{SLS00}), may be impossible if the magnetic
field penetrates regions of the core where neutrons are superfluid
and proton superconducting.

Any observed magnetic field, either based on crustal or core currents, 
has to penetrate the crust matter, thereby affecting its transport
properties. 
Moreover, the structure of the field in the shallow spherical shell layer below the 
surface has a direct  effect on the surface temperature and leads to
a non-uniform distribution with observational consequences 
(Page \cite{P95}) if its strength is above $10^{10}$ G.
Roughly, the effects of the magnetic field onto the transport 
processes can be divided into classical and quantum ones (see, e.g., 
Yakovlev \& Kaminker \cite{YK94} for a review).
The classical effects are due to the Larmor rotation of the electrons,
the main carriers of charge and heat, and are determined by the 
magnetization parameter $\omega_{\scriptscriptstyle B} \tau$ where 
$\omega_{\scriptscriptstyle B} \equiv e B/m_e^* c$ is the gyrofrequency
of the electrons, $\tau$ being their relaxation time and $m_e^*$ their effective mass.
Quantized motion of the electrons transverse to the magnetic field
causes the quantum effects which are of importance only if few Landau 
levels are occupied, requiring thus densities
$\rho < \rho_B = 7045 (B/10^{12} \mathrm{G})^{3/2} \frac{<A>}{<Z>}$ 
g cm$^{-3}$ and temperatures
$T \ll T_B \approx 1.34 \cdot 10^8 (B/10^{12} \mathrm{G})$ K
(Potekhin et al. \cite{PYCG03}) where $<A>$ and $<Z>$ are the average 
mass and charge numbers of the ions.
From these numbers it becomes clear that quantum effects will play an 
important role for  strong magnetic fields in the outermost layers of 
the neutron star crust - the thin low density shell of the envelope - 
but are negligible in the deeper layers.
The thinness of this envelope justifies a study of heat transport in a 
plane parallel, one dimensional approximation and many such 
calculations have been performed (see, e.g., for the most recent one, 
Potekhin \& Yakovlev \cite{PY01}, hereafter PY01, and references therein).
Two-dimensional calculations of heat transport with magnetic 
field have been presented by Schaaf (\cite{S90a},\cite{S90b}) who 
however restricted himself to the thin envelope 
and an uniform magnetic field.
Tsuruta (\cite{T98}) has presented results of two-dimensional cooling 
calculations of neutron stars which included the quantizing effect of 
a dipolar magnetic field in the envelope.
These 2D calculations showed that the 1D approximation is indeed very 
good when the field affects heat transport only in the thin envelope.

The deeper layers of the crust can, however, be affected by classical 
effects in case of strong fields, i.e., large $\omega_{\scriptscriptstyle B}$, 
and/or low temperature, i.e., high $\tau$.
In such a case the usual assumption of an isothermal crust is 
questionable and this is the issue we want to address in this paper.
In particular, in the case of a crustal magnetic field its strength in the 
crust unavoidably exceeds its surface value by one to two orders of magnitude
(see, e.g., Page et al. \cite{PGZ00}) \
and its effect can naturally be expected to be stronger than in the case
of a field permeating the whole star.
This may open a way to study the structure of the magnetic field in 
the crust and provide observational features which will allow us to 
discriminate between the above mentioned two types of hypothesized 
field structure, crustal vs. core.

The paper is organized as follows: 
In the next Sect. ~\ref{sec:input} the basic equations are introduced which 
describe the magnetic field and the heat transport influenced by it. 
The components of the heat conductivity tensor are given and the outer 
boundary condition is discussed. 
The physical input as well as the numerical method are shortly described. 
In Sect. \ref{sec:results} the results of the  numerical calculations are presented. 
For different core temperatures, magnetic field strengths and geometries the 
crustal temperature profiles, the surface temperature distributions and the 
corresponding luminosities are calculated.
Sect.~\ref{sec:discon} is devoted to the discussion of the consequences of the 
magnetic field effects on to the crustal and surface temperature distribution.

\section{Physics Input}
\label{sec:input}

\subsection{Heat transport and magnetic field}
\label{sec:magn-param}

The thermal evolution of the crust is determined by 
the energy balance equation
\be
C\,\frac{\partial T}{\partial t} = 
 \mathrm{Sources} - \mathrm{Sinks} - \vec{\nabla} \cdot \vec{F} 
\label{equ:energy_balance}
\ee
and the heat transport equation
\be
\vec{F} = - \hat{\kappa} \cdot   \vec{\nabla} T
\label{equ:heat_transport}
\ee
where $C$ denotes the specific heat per unit volume,
$\vec{F}$ is the heat flux density and $\hat{\kappa}$ the tensor of 
heat conductivity.

In this paper we intend to consider only the effect of the crustal 
magnetic field on to the stationary temperature distribution in the 
crust, which, in a first approximation, is assumed to be free of 
heat sources and sinks.
The cooling process itself as well as the back reaction of the now 
non-spherically symmetric temperature distribution on to the magnetic 
field decay will be beyond the scope of this work.

\begin{figure*}[ht]
   \centering
   \resizebox{\hsize}{!}{\includegraphics{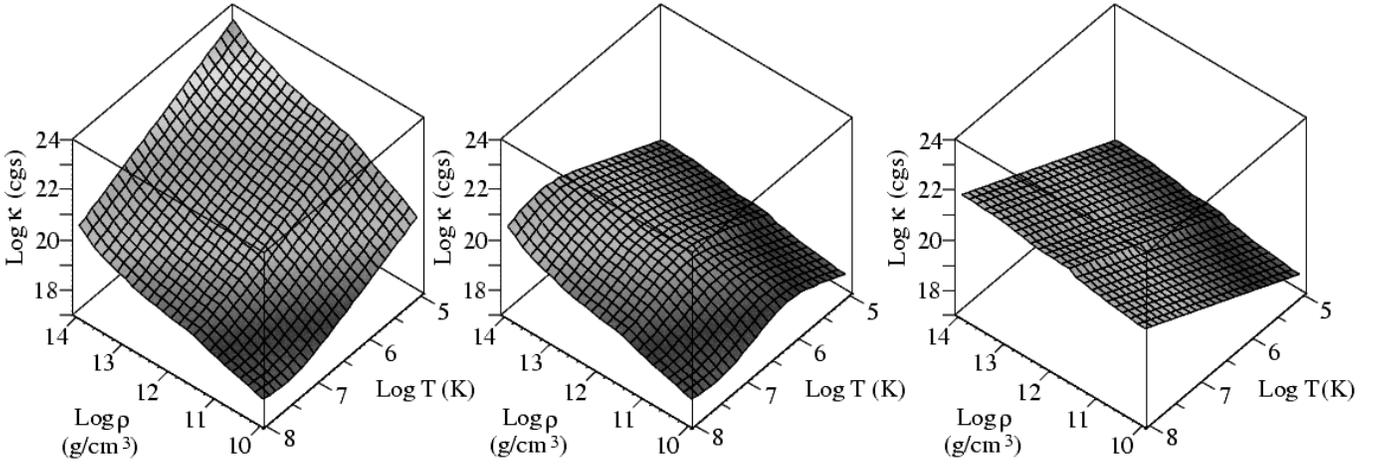}}
   \caption{Thermal conductivity $\kappa_0$, in cgs units, versus density $\rho$ 
            and temperature $T$.
            The left panel shows the phonon-only contribution $\kappa_{0 \; \rm ph}$, 
            the right one the impurity-only contribution $\kappa_{0 \; \rm imp}$, and
            the central panel the complete $\kappa_0$ 
            (with an impurity concentration $Q_\mathrm{imp} = 0.1$).}
   \label{fig:conduct}
\end{figure*}

In the relaxation time approximation the components of $\hat{\kappa}$ 
parallel and perpendicular to the magnetic field, $\kappa_\|$ and 
$\kappa_\perp$ respectively, as well as the Hall component,
$\kappa_\wedge$, are related to the scalar heat  conductivity 
$\kappa_0$ and to the magnetization parameter 
$\omega_{\scriptscriptstyle B} \tau$ by (Yakovlev \& Kaminker \cite{YK94})
\be
\kappa_\| = \kappa_0 \,,\quad  
\kappa_\perp = \frac{\kappa_0}{1+(\omega_{\scriptscriptstyle B} \tau)^2} \,, \quad
\kappa_{\wedge} =  \omega_{\scriptscriptstyle B} \tau \, \kappa_\perp 
\label{kappa}
\ee
For $\kappa_0$ we use
\be
\kappa_0 = \frac{\pi^2 k_B^2 T n_e}{3 m_* \nu}
\label{eq:kappa_0}
\ee
where $\nu = 1/\tau$ is the effective electron collisional frequency and is given
by the sum
\be
\nu = \nu_\mathrm{ph} + \nu_\mathrm{ion} + \nu_\mathrm{imp}
\ee
where 
$\nu_\mathrm{ph}$ is the effective collisional frequency for electron-phonon,
$\nu_\mathrm{ion}$ for electron-ion, and
$\nu_\mathrm{imp}$ for electron-impurity collisions.
Since for the range of densities and temperatures we are interested in the
crust is always in the solid state we neglect $\nu_\mathrm{ion}$
and for $\nu_\mathrm{ph}$ we use the calculations Baiko \& Yakovlev (\cite{BY96})
while we take $\nu_\mathrm{imp}$ from  Yakovlev \& Urpin (\cite{YU80}).
Everywhere in this work we assume an ``impurity parameter'' 
$Q \equiv x_\mathrm{imp} \overline{(\Delta Z)^2}$ equal to 0.1,
where $x_\mathrm{imp}$ is the fractional number of impurities which have
a mean square charge excess of $\overline{(\Delta Z)^2}$.
Figure~\ref{fig:conduct} shows the resulting $\kappa_0$:
note that $\nu_\mathrm{imp}$ is almost temperature independent while
$\nu_\mathrm{ph}$ scale approximately as $T^2$ so that if $\nu = \nu_\mathrm{ph}$
then the ``phonon-only conductivity'' $\kappa_{0 \; \mathrm{ph}}$ $\sim T^{-1}$ 
(Fig.~\ref{fig:conduct} left panel) but if $\nu = \nu_\mathrm{imp}$ we would have 
an ``impurity-only conductivity'' $\kappa_{0 \; \mathrm{imp}}$ $\sim T^{+1}$ 
(Fig.~\ref{fig:conduct} right panel) and the  exact $\kappa_0$ (Fig.~\ref{fig:conduct} 
central panel), which one could write as 
$\kappa_0 = (1/\kappa_{0 \; \mathrm{ph}} +1/\kappa_{0 \mathrm{imp}})^{-1}$,
shows both types of
behaviours depending on which type of collision process dominates.

\begin{figure}[hb]
   \centering
   \resizebox{\hsize}{!}{\includegraphics{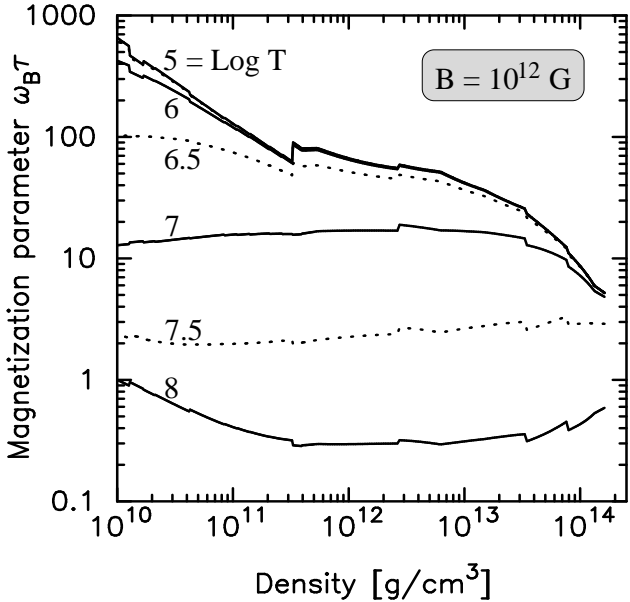}}
   \caption{Magnetization parameter $\omega_{\scriptscriptstyle B} \tau$ 
            vs. density at six different temperatures (as labeled on
            the curves) assuming a uniform magnetic field 
            of strength $B =10^{12}$ G.
            Its value for different field strengths scales 
	    linearly in $B$.}
   \label{fig:magnetization}
\end{figure}

\begin{figure}[hb]
   \centering
   \resizebox{\hsize}{!}{\includegraphics{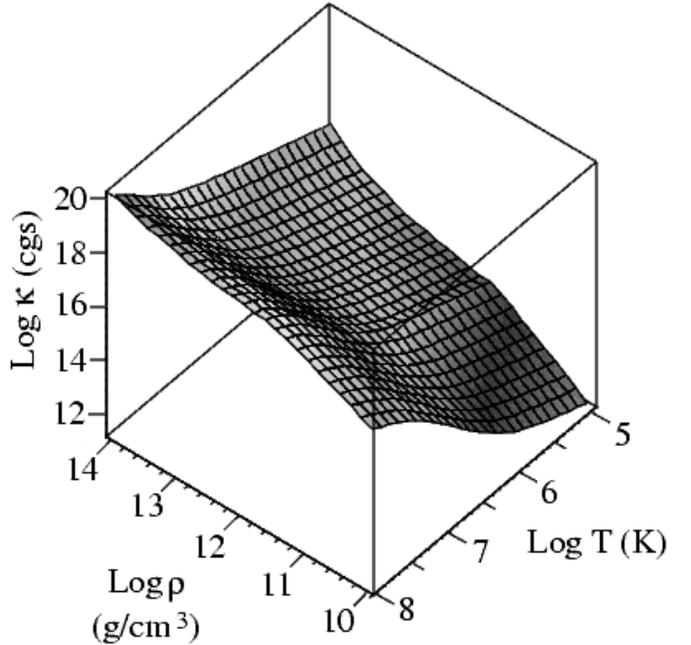}}
   \caption{Thermal conductivity, in cgs units, perpendicular to the magnetic field,
            $\kappa_\perp$, vs. density $\rho$ and temperature $T$
            for a uniform magnetic field of strength $3\times 10^{12}$ G.
            For field strengths $\gg 10^{12}$ G, i.e., for 
            $\omega_{\scriptscriptstyle B} \tau \gg 1$,
            $\kappa_\perp$ scales as $B^{-2}$.
            The impurity parameter $Q$ is assumed to be 0.1.}
   \label{fig:kappa_perp}
\end{figure}

The magnetization parameter $\omega_{\scriptscriptstyle B} \tau$ varies 
strongly throughout the crust by many orders of magnitude. 
In $\omega_{\scriptscriptstyle B}$ both $B$ and $m_e^*$ span a large range of 
values throughout the crust and in $\tau$ there is also a strong dependence 
on the temperature $T$, chemical composition and the thermodynamic 
phase of the matter. 
For illustration we show, in Fig.~\ref{fig:magnetization},
$\omega_{\scriptscriptstyle B} \tau$
in the crust at various {\em uniform} temperatures and a 
{\em uniform} field strength: 
however,  neither $T$ nor $B$ will be uniform in our realistic 
calculation presented below.
Note that values of $\omega_{\scriptscriptstyle B} \tau$ at $T= 10^5$ and $10^6$ K
are very close to each other because at such low $T$ $\nu$ is dominated by
$\nu_\mathrm{imp}$ which is temperature independent while when going to
increasingly higher temperatures $\nu_\mathrm{ph}$ contributes more and more
and hence $\tau$ decreases.
Finally, Fig.~\ref{fig:kappa_perp} shows the resulting $\kappa_\perp$':
notice its very different $T$--$\rho$ behaviour compared to $\kappa_0$
due to the simple fact that at high $\omega_{\scriptscriptstyle B} \tau$
we obtain $\kappa_\perp \propto \tau^{-1}$ while $\kappa_0 \propto \tau$.

\subsection{The Crustal Magnetic Field}

A dipolar poloidal magnetic field can be conveniently described in 
terms of the (possibly time dependent) Stoke's stream function 
$S = S(r,t)$.
The vector potential $\vec{A} = (0, 0, A_{\varphi})$ is written as
$A_{\varphi} = S(r, t) \sin \theta /r$, where $r$, $\theta$ and 
$\varphi$ are spherical coordinates. 
The field components are then expressed as
\be
B_r      = + \frac{2 \cos \theta}{r^2} S(r,t) 
         = B_0 \; \frac{\cos \theta}{x^2} \; s(x,t)
\label{equ:Stokes-r}
\ee
\be
B_\theta = - \frac{\sin \theta}{r} 
             \frac{\partial S(r,t)}{\partial r}
         = - \frac{B_0}{2} \; \frac{\sin \theta}{x}  \;
             \frac{\partial s(x,t)}{\partial x}
\label{equ:Stokes-theta} 
\ee
\be
B_\varphi = 0
\ee
where $B_0$ is the field strength at the magnetic pole, and we have
introduced normalized variables $S=s B_0 R_{N\!S}^2/2$, with $s=1$ 
at the magnetic north pole, and $x=r/R_{N\!S}$.
The vacuum solution, outside the star, is simply $s =1/x$.
At the surface, $x=1$, the standard boundary condition for 
matching to a vacuum dipole field should be fulfilled,
\be
\frac{\partial S}{\partial r} + \frac{S}{R_{N\!S}} = 0
\;\;\;\;\; \mathrm{i.e.} \;\;\;\;\;
\left. \frac{\partial s}{\partial x}\right|_{x=1} = -s
\ee
and at the center regularity requires $S(r=0,t) \equiv 0$.

For the present investigation we select for the crustal field 
configuration a snapshot of the evolution of $S(x,t)$. 
The Stoke's function $S(x,t)$ was calculated by solving the induction 
equation, applying the above boundary 
conditions and an electric conductivity which reflects the same 
microphysics as the heat conductivity does for the model under consideration 
(see, e.g., Geppert \& Urpin~\cite{GU94} or Page et al. \cite{PGZ00}).

\begin{figure}
   \centering
   \resizebox{\hsize}{!}{\includegraphics[width=7.0cm]{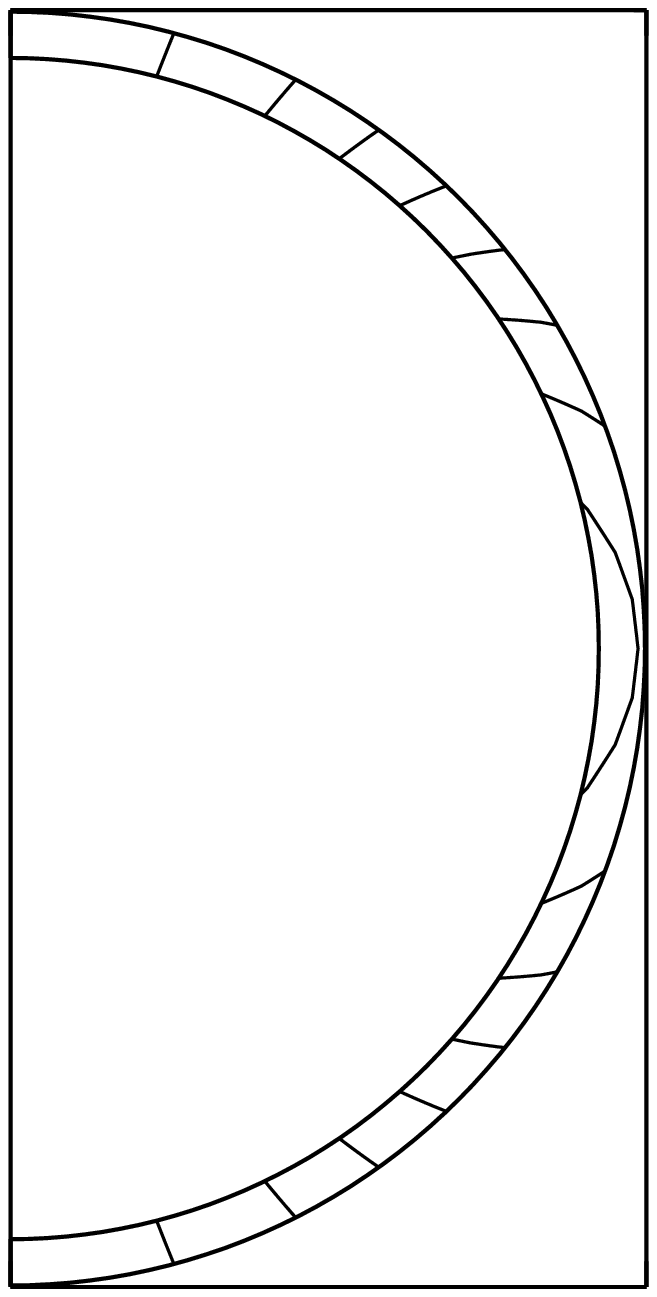}
                         \includegraphics[width=7.0cm]{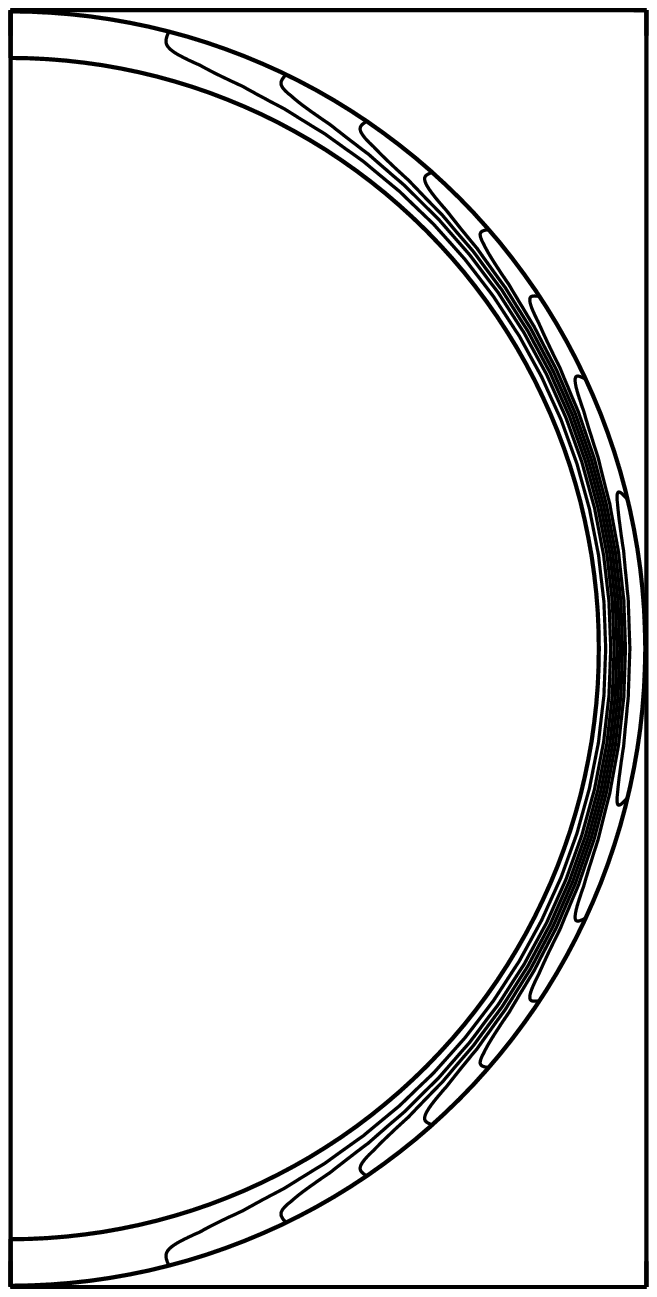}}
   \caption{Magnetic field lines of the core field 
            (left panel) and of the crustal field as applied in our
	    calculations. Both field configurations match for $r \ge R_{NS}$
	    with a dipolar poloidal vacuum field. They are presented 
	    in a meridional plane,  of the crust  
	    for $\rho_n \ge \rho \ge \rho_b$ (see Sects. ~\ref{sec:bound} and
	    ~\ref{sec:EOS-chem}). 
            The polar surface value $B_0$ is identical for both field
            types.}
   \label{fig:field_types}
\end{figure}

The initial value $S(x,t=0)$ is {\em a priori} unknown and depends 
on the field generation process. 
For a crustal magnetic field, the function $S(x,t=0)$ initially 
vanishes in  the core and, due to proton superconductivity, the 
Meissner-Ochsenfeld effect prevents the field from penetrating into 
the core (see, e.g. Page et al. \cite{PGZ00}). 
A typical crustal field structure is shown in the right panel of 
Fig.~\ref{fig:field_types}. 
In case the magnetic field is maintained by electric currents 
circulating in the core 
the field penetrates the crust too but has a 
qualitatively different structure. 
Let us assume that for the core field there are no currents in the crust 
and the field is maintained in the core by axisymmetric currents 
circulating around the center of the star. 
Then a dipolar field will penetrate the crust with the components
\be
B_r^{dipole}      =  B_0 \; \frac{\cos \theta}{x^3} \; ,
\ee
\be
B_\theta^{dipole} =  \frac{B_0}{2} \; \frac{\sin \theta}{x^3} \; ,
\ee
\be
B_\varphi^{dipole} = 0 \; ,
\ee
where $B_0$ denotes again the polar surface value of the magnetic field.
The crustal field lines of such a field are presented in the left
panel of Fig.~\ref{fig:field_types}. 
A comparison of the geometries of
both fields immediately suggests that the star-centered core field will not 
significantly affect the heat transport through the crust while a 
crustal field may cause drastic changes.
Moreover, in the crust $B_\theta$ has to be much larger than $B_\theta^{dipole}$,
for a given identital external field, since the flux of a crustal field is compressed into 
a layer of thicknes $\Delta r \sim \frac{1}{10} - \frac{1}{20}$ of $R_{N\!S}$,
while the flux of a core field can expand within the whole core,
i.e., typically $B_\theta \sim 10 - 20 B_\theta^{dipole}$.

\subsection{The Two-Dimensional Heat Transport Equation}

Denoting the components of the temperature gradient $\vec{\nabla}T$
parallel and perpendicular to the unit vector of the magnetic field, 
$\vec b$ as well as the Hall, component by
\ba
(\vec{\nabla} T)_\|     = \vec{b} \; (\vec{\nabla} T \cdot \vec{b}) 
                   \;\; &,& \;\;
(\vec{\nabla} T)_\perp  = \vec{b} \times (\vec{\nabla} T \times \vec{b})  
\;\; ,
\\
(\vec{\nabla} T)_\wedge &=& \vec{b} \times \vec{\nabla} T
\;\; ,
\nonumber
\ea
the magnetic--field--influenced heat flux can be expressed by  
\ba
\hat{\kappa} \cdot \vec{\nabla}T =     
\nonumber \\
\kappa_{\|}     \; \vec{b} \; (\vec{\nabla} T \cdot \vec{b}) +
\kappa_{\perp}  \; \vec{b} \times (\vec{\nabla} T \times \vec{b}) +
\kappa_{\wedge} \; \vec{b} \times \vec{\nabla} T
\;\; .
\ea
This provides the following expression for the heat flux 
in terms of the scalar heat conductivity, 
the magnetization parameter, the temperature gradient and the unit 
vector of the magnetic field:
\ba
\vec{F} = 
- \hat{\kappa} \cdot \vec{\nabla} T =
- \; \frac{\kappa_0}{1+(\omega_{\scriptscriptstyle B}\tau)^2} \times
\nonumber \\
        \left[ \vec{\nabla}T + 
       (\omega_{\scriptscriptstyle B}\tau)^2 \; \vec{b} \; (\vec{\nabla}T \cdot \vec{b}) +
			\omega_{\scriptscriptstyle B}\tau \; \vec{b} \times \vec{\nabla}T
        \right]\,\, .
\ea
The last term on the r.h.s. represents the Hall component of the 
heat flux. Its divergence vanishes as long as the magnetic field and 
the temperature gradient are assumed to be axially symmetric, 
i.e., do not depend on the azimuthal angle $\varphi$. 
By use of the representation of the dipolar magnetic field in terms of
the Stoke's stream function $s$	and its radial derivatives 
(see Eqs.~\ref{equ:Stokes-r}, \ref{equ:Stokes-theta}) and by 
introducing the heat conductivity coefficients
\be
\chi_1= \frac{\kappa_0}{1+(\omega_{\scriptscriptstyle B}\tau)^2}\,\,,\,\,\chi_2=
	\frac{\kappa_0\,(\omega_{\scriptscriptstyle B_0}\tau)^2}{1+(\omega_{\scriptscriptstyle B}\tau)^2}\,\,,
\label{equ:chi}
\ee
the radial and meridional components of the magnetic--field--dependent
heat flux have the following form:		
\ba
F_x= -\chi_1\frac{\partial T}{\partial x} + 
\nonumber \\
      \chi_2 \left( 
      \frac{\partial T}{\partial \theta} \frac{s}{2x^4}
       \frac{\partial s}{\partial x}
         \sin{\theta}\cos{\theta} - \frac{\partial T}{\partial x} 
         \frac{s^2}{x^4} \cos^2{\theta}\right)\,\,,
\label{equ:F_r-chi}
\ea
\ba
F_{\theta}= -\chi_1\frac{1}{x} \frac{\partial T}{\partial \theta} + 
\nonumber \\
             \chi_2 \left( 
              \frac{\partial T}{\partial x} \frac{s}{2x^3}
              \frac{\partial s}{\partial x}
      \sin{\theta} \cos{\theta} - \frac{\partial T}{\partial \theta} 
        \frac{1}{4x^3}
       \left(\frac{\partial s}{\partial x}\right)^2 
       \sin^2{\theta}\right)\,\,,
\label{equ:F_t-chi}
\ea
where $T$ is measured in units of $T_\mathrm{core}$ and the heat flux is
normalized on $T_\mathrm{core}/R_{N\!S}$.
Note that in Eq.~\ref{equ:chi} the magnetization parameter 
$\omega_{\scriptscriptstyle B_0}\tau$ is calculated by use of the polar surface
field strength $B_0$.

Axial symmetry implies that, along the polar axis, $F_\theta$ and 
$\partial F_\theta/\partial \theta$ vanish while $x^2 F_x$ is 
constant.
Mirror symmetry at the magnetic equator implies that $F_\theta$ 
vanishes (but $\partial F_\theta/\partial \theta$ can be very large).

Our aim in this paper will be to find stationary solutions for the 
temperature distribution in the neutron star crust, i.e. to solve the 
equation
\be
\vec{\nabla} \cdot \vec{F} = 
       \frac{1}{x^2} \frac{\partial (x^2 F_x)}{\partial x} +
       \frac{1}{x \sin \theta} 
  \frac{\partial (\sin \theta F_\theta)}{\partial \theta} = 0 \;\; .
\ee
The boundary condition for this equation at the crust-core boundary
is `` $T$ fixed and uniform '' and at the surface it as discussed in the
following subsection.

\subsection{Outer boundary: $T_b(\vec B)-T_s(\vec B)$ relationship
            \label{sec:bound}}

In the lowest--density layers, close to the surface, matter is no 
longer degenerate and the magnetic field affects the equation of 
state. 
Appropriate treatment of this layer requires solving for hydrostatic 
equilibrium simultaneously with heat transport.
To avoid this problem, we separate this layer, called envelope, 
from the crust and incorporate it in the outer boundary condition. 
We stop the integration at an outer boundary density $\rho_b$, at 
radius $r_b$, and, for each latitude $\theta$, obtain a radial flux 
$F_r(\rho_b, \theta)$ and a temperature 
$T(\rho_b, \theta)$ 
which we match to the envelope.
In the envelope approximation, a surface temperature $T_s$, and hence 
an out coming flux $F^\mathrm{env} \equiv \sigma T_s^4$, is chosen and 
hydrostatic  equilibrium and heat transport are solved toward 
increasing densities up to $\rho_b$ giving the temperature 
$T^\mathrm{env}_b$ at that density. 
Varying $T_s$ gives a ``$T_b - T_s$ relationship''.
The matching of the envelope with our interior calculation is simply 
obtained by imposing 
$T^\mathrm{env}_b = T(\rho_b, \theta)$
and 
$F^\mathrm{env} \equiv \sigma T_s^4 = F_r(\rho_b, \theta)$
which is our outer boundary condition.
The magnetic field strongly affects the structure and transport 
properties of the envelope and the ``$T_b - T_s$  relationship'': 
at each point $(r_b,\theta)$ we apply an envelope with a magnetic field 
$\vec{B}$ equal to the field we have at that point,
hence giving us a $T_b(\vec B)-T_s(\vec B)$ relationship.

Many calculations of magnetized envelopes have been presented and we 
use the latest and most accurate one from PY01.
These authors pose the bottom of the envelope at the neutron drip 
density, i.e., $\rho_b \approx 4\cdot 10^{11}$ g cm$^{-3}$, while we
intend to apply lower values down to $\rho_b = 10^{10}$ g cm$^{-3}$ 
in order to extend the 2--dimensional transport calculation as far as 
possible while still being safely in the non-quantizing regime.
Since the temperature profile in the 1--dimensional envelope 
calculations  of Potekhin \& Yakovlev is quite flat between 
$4\cdot 10^{11}$ g cm$^{-3}$ and $10^{10}$ g cm$^{-3}$ the same 
$T_b - T_s$ relationship can be applied for the lower $\rho_b$ which we 
consider a good approximation.
Explicitly, we apply their equations (26) to (30) but replace their 
$T_\mathrm{core}$ which they assume to be spherically symmetric by our
calculated angle-dependent $T_b(\theta)$. 

To illustrate the main features of this boundary condition, a good 
approximation (Greenstein \& Hartke \cite{GH83};YP01) is to write it as 
\be
F_r(\rho_b,\Theta_B,B,T_b) \approx
F_\| \cos^2 \Theta_B + F_\perp  \sin^2 \Theta_B \,\,,
\label{equ:boundary}
\ee
which expresses $F_r$, for an arbitrary angle $\Theta_B$ between
$\vec{B}$ and the radial direction, in terms of the radial flux for a
radial field $F_\| \equiv F_r(\rho_b,\Theta_B=0,B,T_b)$
and for a meridional field $F_\perp \equiv F_r(\rho_b,\Theta_B=\pi/2,B,T_b)$.
For a dipolar field $\Theta_B$ is related to the polar angle $\theta$  by
$\tan \Theta_B = 0.5 \tan \theta$.
Note that when $B \gg 10^{11}$~G one has $F_\perp \ll F_\|$,
i.e., the envelope is blanketing the star much more strongly
around the magnetic equator than around the poles.

\subsection{Equation of state and chemical composition}
            \label{sec:EOS-chem}

We will consider a $1.4 M_{\odot}$ neutron star whose structure 
is obtained by integrating the Tolman-Oppenheimer-Volkoff equation
of hydrostatic equilibrium.
The core matter is described by the equation of state (EOS) 
calculated by Wiringa et al. (\cite{WFF88}).
We separate the crust from the core at the density 
$\rho = \rho_{n} \equiv 1.62 \times 10^{14}$ g cm$^{-3}$
(Lorenz et al. \cite{LRP93})
and use the EOS of Negele \& Vautherin (\cite{NV73}) for the inner crust,
at $\rho > \rho_{drip} \equiv 4.4 \times 10^{11}$ g cm$^{-3}$,
and Haensel et al. (\cite{HZD89}) for the outer crust.
We assume that the chemical composition is that of cold catalyzed matter,
as is the case for these two crustal EOSs.

The star so produced has a radius $R_{NS} = 10.9$ km and the 
crust-core boundary is at radius $r_{n} = 0.925 R_{NS} = 9.33$ km.

\subsection{Numerical method}

\begin{figure*}[ht]
   \centering
   \includegraphics[width=18.3cm]{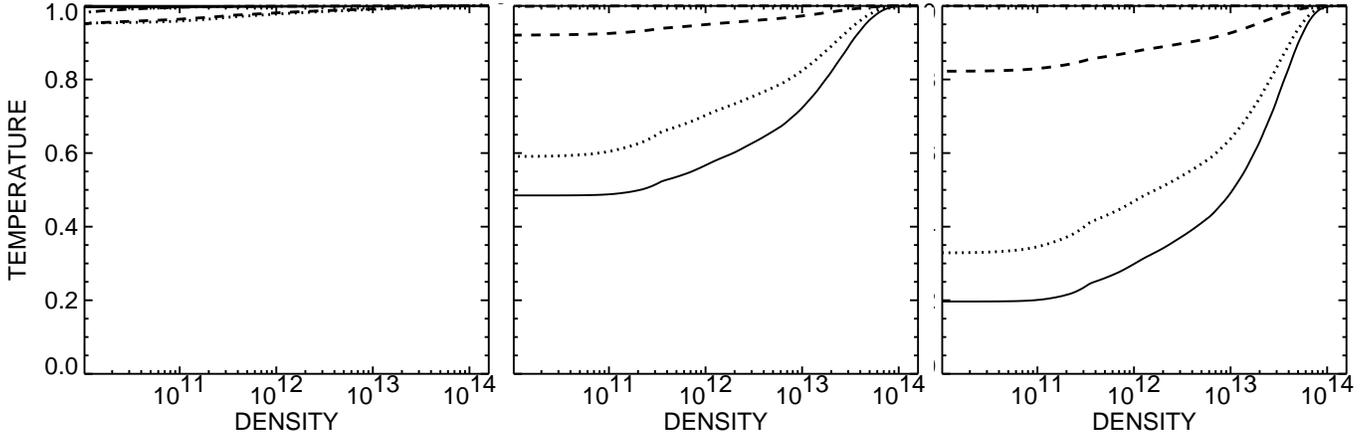}
   \caption{The crustal temperature $T$ (normalized on $T_\mathrm{core}$) as 
            a function of the density within the crust at 4 different
            polar angles: 
            $\theta=0$: dotted-dashed line (almost coincident with the upper 
            limit of the figure), 
            $\theta=30$: dashed line, 
            $\theta=60^0$: dotted line, and 
            $\theta=90^0$: full line, 
            for a crustal field of strength $B_0=3\cdot 10^{12}$G.
            The three panels correspond, from left to right, to
	    $T_\mathrm{core}=10^8$K, $10^7$K, and $10^6$K, respectively.}
   \label{fig:b3e12_cru_dens}
\end{figure*}

We solve the heat transport equation in its time-dependent form,
\be
 C \frac{\partial T}{\partial t} = - \vec{\nabla} \cdot \vec{F}
\ee
starting with constant temperature, $T = T_\mathrm{core}$. As boundary condition 
at the lower boundary the temperature is kept fixed to the value 
$T_\mathrm{core}$ at all times while at the outer boundary the radial heat flux 
is related to the temperature at the same point as described above. 
The code is then run until a stationary state is reached.

For the discretization of the spatial derivatives a staggered mesh 
method as described in Stone \& Norman (\cite{SN92}) is used in 
spherical polar coordinates ($r$ and $\theta$). 
For the integration in time the fully explicit method turned out to 
be prohibitively slow because the diffusion coefficients vary 
strongly in space in the case of strong magnetic fields. 
We therefore use the scheme in an operator-split implementation that 
treats the radial diffusion terms implicitly while the horizontal 
diffusion and the $ {\partial T}/\partial \theta $ term in $F_r$ 
remain explicit. 
This keeps the computational effort per time step small as the 
equations to be solved are tridiagonal, but allows for a sufficiently 
large time step to reach a stationary state within several hours of 
computing time. 
For the evaluation of the transport coefficients and the outer 
boundary condition the temperature resulting from the last time step 
is always used, i.e., only the radial derivatives are treated 
implicitly.

\section{Results}
\label{sec:results}
 
\subsection{Internal temperature}

\begin{figure}
   \centering
   \includegraphics[width=6.2cm]{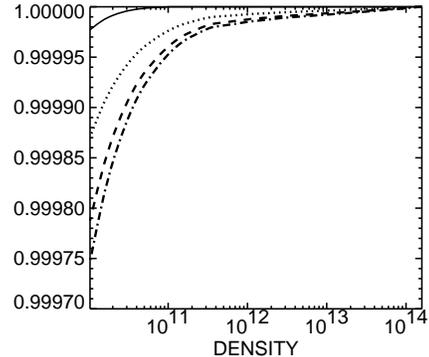}
   \caption{Same as Fig.~\protect\ref{fig:b3e12_cru_dens} but for
            a dipolar core field with $T_\mathrm{core}=10^7$K.}
   \label{fig:b3e12_10_8_dip_dens}
\end{figure}

\begin{figure}[h]
 \resizebox{\hsize}{!}{\includegraphics[width=7.0cm]{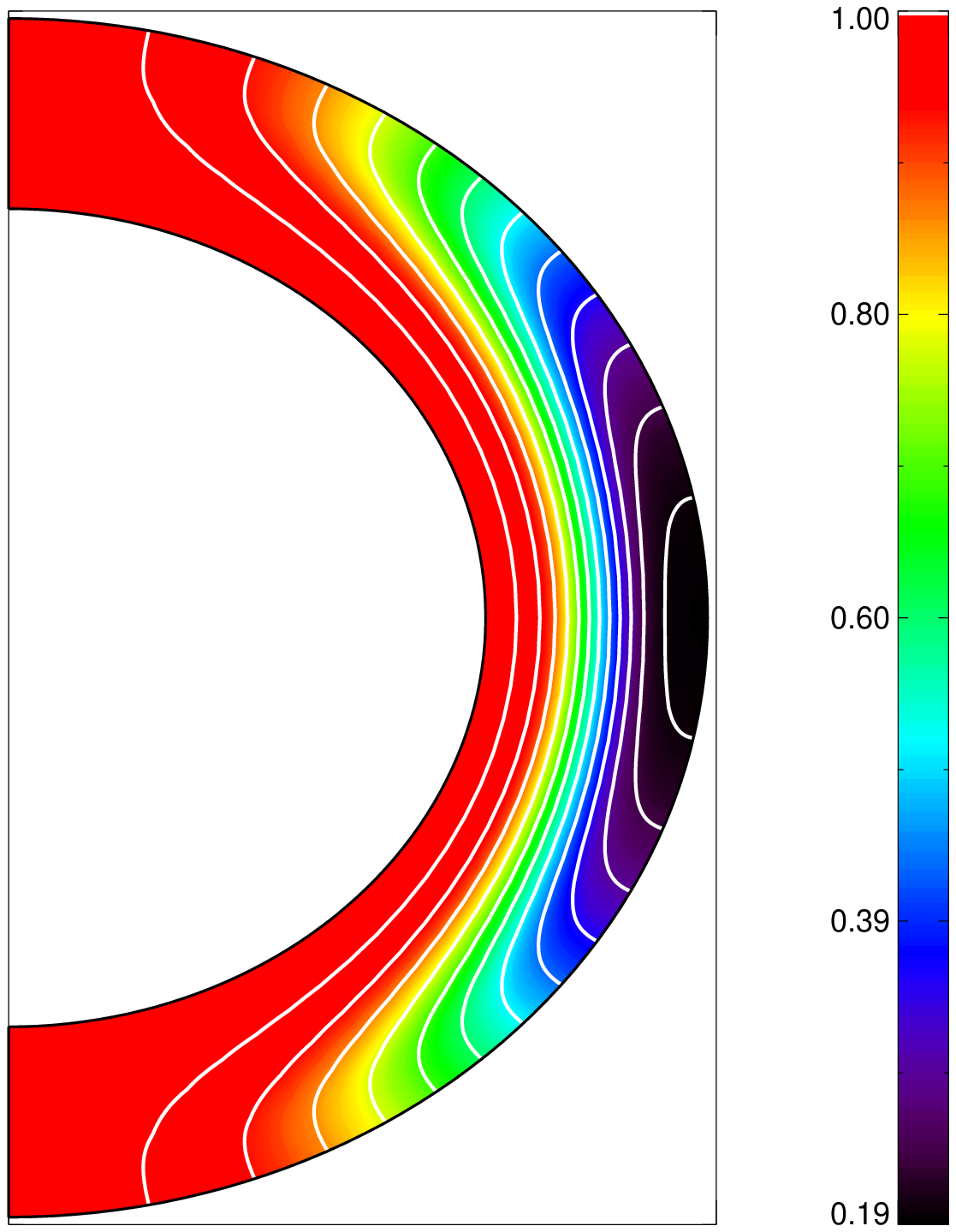}
                       \includegraphics[width=7.0cm]{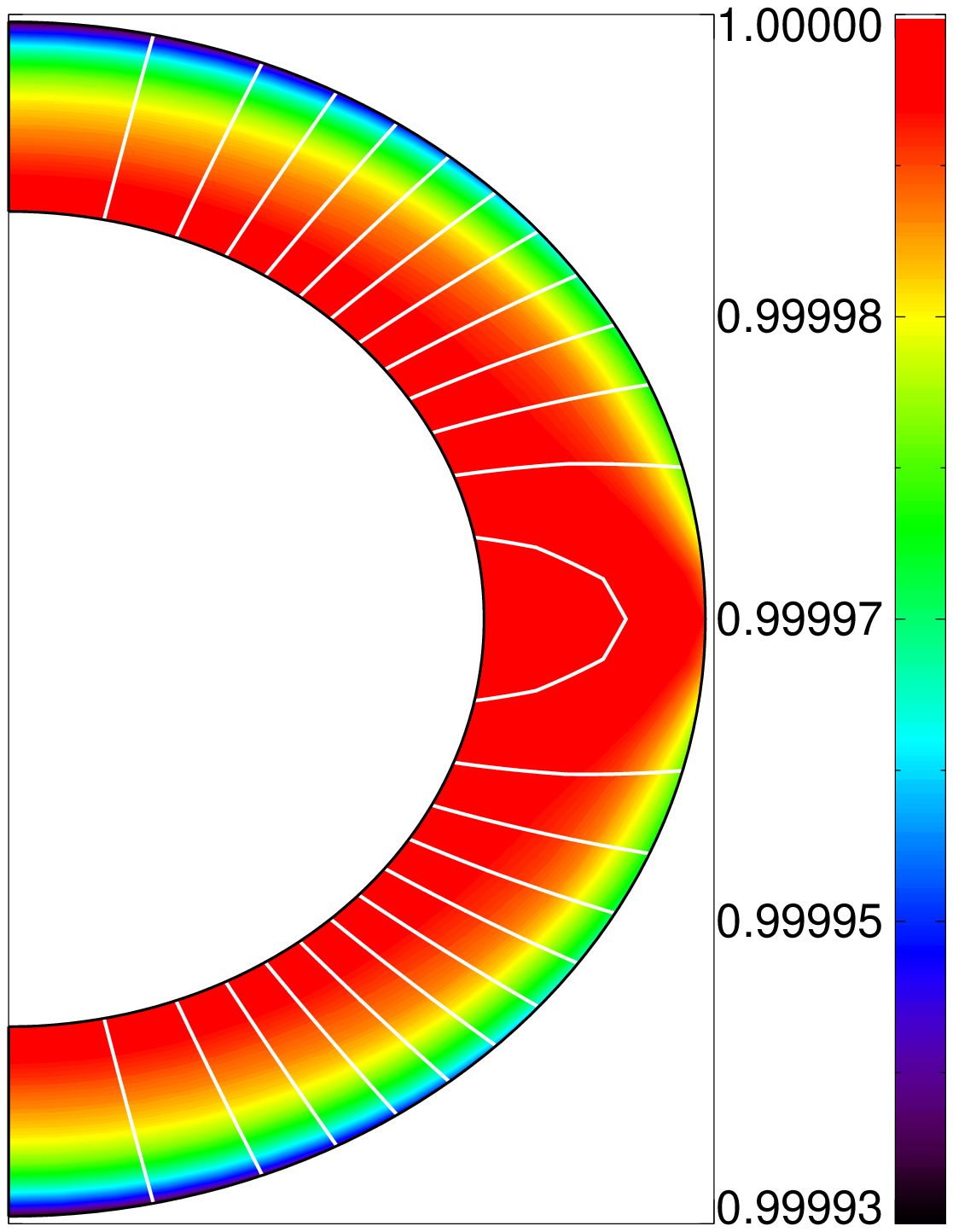}}
 \caption{Representation of both field lines and temperature 
          distribution in the  crust whose radial scale 
          ($r(\rho_n) \le r \le r(\rho_b)$) is stretched by a factor 
          of $5$, assuming $B_0=3\cdot 10^{12}$G and $T_\mathrm{core}=10^6$K. 
          Left panel corresponds to a crustal field, 
          right panel to a star-centered core field.
          Bars show the temperature scales in units of $T_\mathrm{core}$.}
 \label{fig:heat_field}
\end{figure}

\begin{figure}
   \centering
   \resizebox{\hsize}{!}{\includegraphics{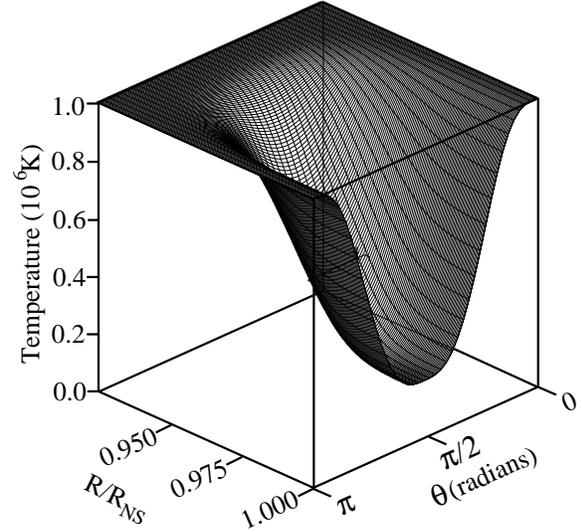}}
   \caption{3-D presentation of the temperature distribution in the
            crust for $T_\mathrm{core} = 10^6$ K and a crustal field
            with strength $B_0 = 3\cdot 10^{12}$ G.
           (corresponding to the right panel of Fig.~\protect\ref{fig:b3e12_cru_dens}).}
   \label{fig:3D_b3e12_t6_rho10}
\end{figure}

The importance of the magnetic-field-induced non-isothermality of the 
whole crust is illustrated in Figs.~\ref{fig:b3e12_cru_dens}, 
\ref{fig:b3e12_10_8_dip_dens}, \ref{fig:heat_field}, and 
\ref{fig:3D_b3e12_t6_rho10}.
The temperature profiles of Fig.~\ref{fig:b3e12_cru_dens}
when compared with Fig.~\ref{fig:b3e12_10_8_dip_dens},
show clearly the difference between a crustal and a core field,
the latter inducing temperature variations in the crust of much less 
than 1\% at $B_0 = 3\cdot 10^{12}$ G while the former can result 
in variations of a factor two for the same dipolar external field 
strength and $T_\mathrm{core}= 10^7$K, and even much larger at lower $T_\mathrm{core}$.

For strong fields, when $\omega_{\scriptscriptstyle B}\tau \gg 1$,
one has $\kappa_{\|} =\kappa_0 \gg \kappa_{\perp}$  and hence heat flows
essentially along the field lines and, given the large values of
$\kappa_0$, no large temperature gradient can build up along them,
as illustrated in Fig.~\ref{fig:heat_field}. 
Only extremely strong core fields
may cause significant deviations from the isothermality of the crust; 
for $B_0 = 10^{16}$G we could observe a difference of only 
$10$ \% for $T(\rho_b)$ between pole and equator.
The qualitative difference between core and crustal magnetic fields is then
easily understood by observing that field lines are essentially
radial in most of the crust in the case of a core field while they are 
predominantly meridional for a crustal field inhibiting radial
heat flow in a large part of the crust.
We could find significant differences to the isothermal crust model only if the 
polar surface field strength exceeds $10^{12}$G. 
Additionally, the core temperature should be smaller than $10^8$K.

In the case of a star-centered core field, 
Fig.~\ref{fig:b3e12_10_8_dip_dens} and the right panel of 
Fig.~\ref{fig:heat_field}
shows that the stellar equator is very slightly warmer than the pole.
This is a direct consequence of the outer boundary condition,
i.e., the envelope: 
magnetic--field--induced anisotropies are weak within the crust for such
fields but are large in the envelope which is more insulating around the
equator than around the poles.

\begin{figure}
 \resizebox{\hsize}{!}{\includegraphics[width=7.0cm]{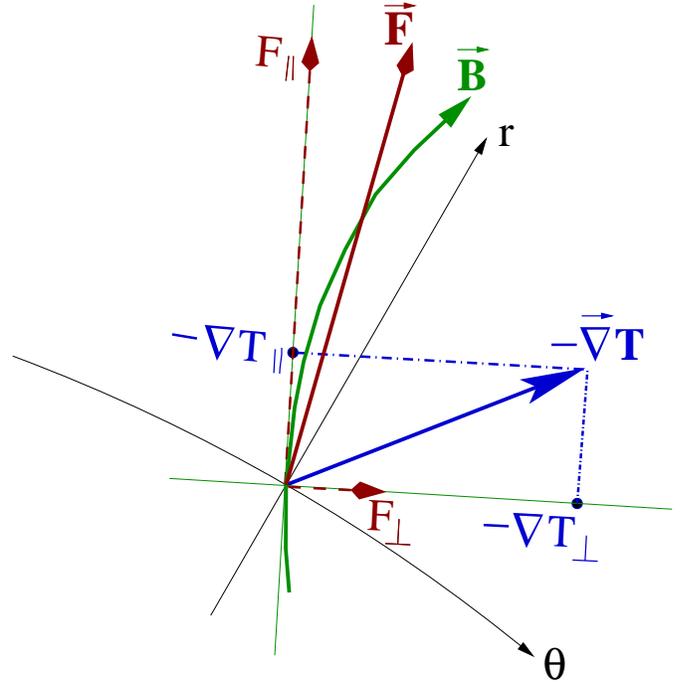}}
 \caption{Illustration of the magnetic-field-induced anisotropy
          (see text for details)}
 \label{fig:drawing}
\end{figure}

A 3D representation of the crust temperature is shown in  
Fig.~\ref{fig:3D_b3e12_t6_rho10}:
it may be surprising in the sense that heat flows into the crust from the
core, at a fixed $T_\mathrm{core}$, and out of the crust at the surface
and most of the heat comes out in the polar regions which are, first,
warmer than the equator and where, second, the envelope is less insulating.
So heat must be flowing from the equatorial regions toward the polar regions,
i.e., apparently from cold regions toward warmer ones !
That this situation cannot violate the second law of thermodynamics
is built into the heat conductivity tensor $\hat{\kappa}$ which is
positive definite and guarantees that $\vec{F} \cdot \vec{\nabla} T < 0$
always.
Figure~\ref{fig:drawing} illustrates this situation: at a point in the northern
hemisphere $\vec{\nabla} T$ is almost perpendicular to $\vec{B}$
and $-(\vec{\nabla} T)_\theta$ is clearly pointing toward the equator.
However, since $F_\| = -\kappa_\| (\vec{\nabla} T)_\|$,  
$F_\perp = -\kappa_\perp (\vec{\nabla} T)_\perp$, and 
$\kappa_\| \gg \kappa_\perp$ we have $|F_\|| \gg |F_\perp|$
and the resulting $F_\theta$ is negative, i.e., pointing toward the pole: 
heat is flowing from the equator toward the poles but does it along 
the magnetic field lines from warmer regions toward colder ones.
Our numerical results show that, e.g., in the northern hemisphere, 
$F_\theta$ is negative in large regions and is positive elsewhere
depending on the orientation of $\vec{B}$.

The results shown in Fig.~\ref{fig:b3e12_cru_dens} for the crustal field
configuration with the typical strength $B_0=3\cdot 10^{12}$G confirms the
statement that with increasing magnetization parameter the anisotropy of the
temperature distribution within the crust increases too. The magnetization
parameter increases with increasing magnetic field strength and with 
decreasing crustal temperature, i.e. with decreasing 
$T_\mathrm{core}$, because the relaxation time of electron--phonon collisions grows
strongly in the course of cooling. Therefore, while the temperature profile
along the poles shows practically no gradient with decreasing $T_\mathrm{core}$
the ratio $T(\rho_b,\theta =90^\circ)/T(\rho_b,\theta = 0^\circ)$ decreases
from 0.95 to 0.5 and 0.2 when $T_\mathrm{core}$ decreases from $10^8$ K to
$10^7$ and $10^6$ K, respectively.
Also, an increase of the magnetic field strength amplifies that difference. 
Applying the same field structure but $B_0=10^{13}$G
the temperature ratio becomes smaller than 0.1 for $T_\mathrm{core}=10^6$K.

Note also that the highly unknown parameter of the impurity concentration
($Q=0.1$ throughout this paper) affects the relaxation time: the more impurities
the shorter the relaxation time of electron--impurity collisions. Therefore, in
a very pure neutron star crust the magnetic field effects onto the crustal
temperature distribution will be even more pronounced.

\subsection{Surface temperature distribution}

\begin{figure*}
   \centering
   \includegraphics[height=8.2cm]{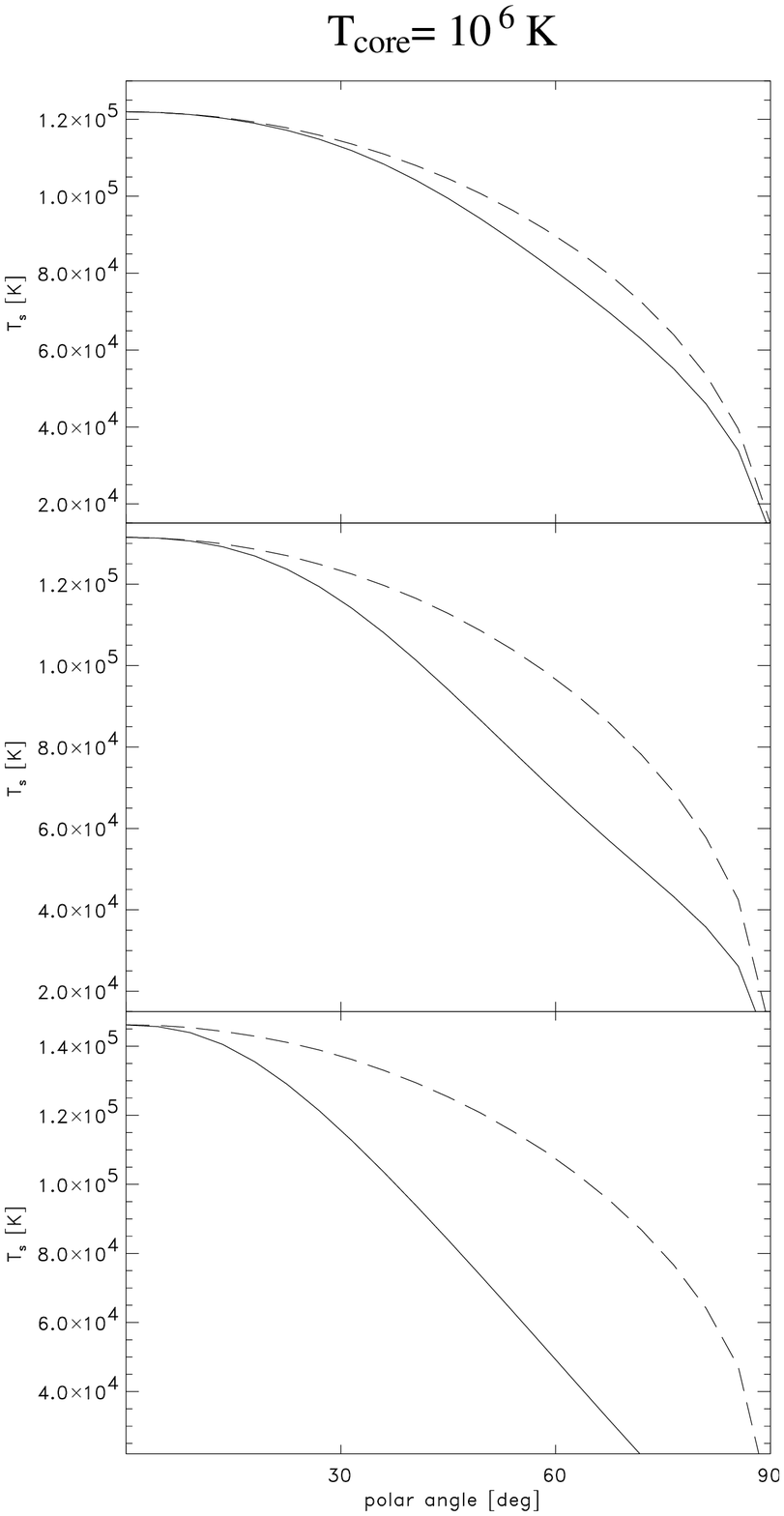}
   \includegraphics[height=8.2cm]{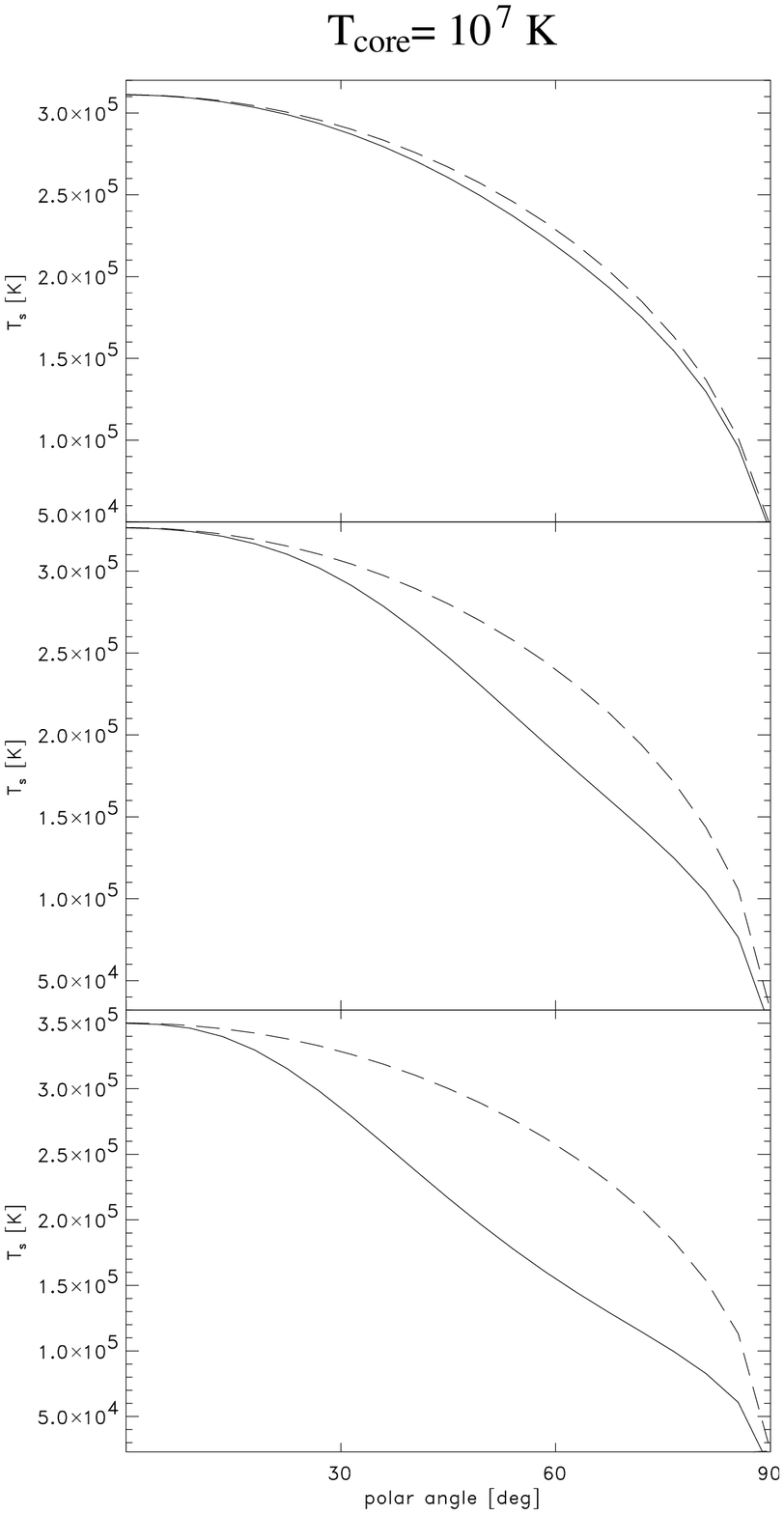}
   \includegraphics[height=8.2cm]{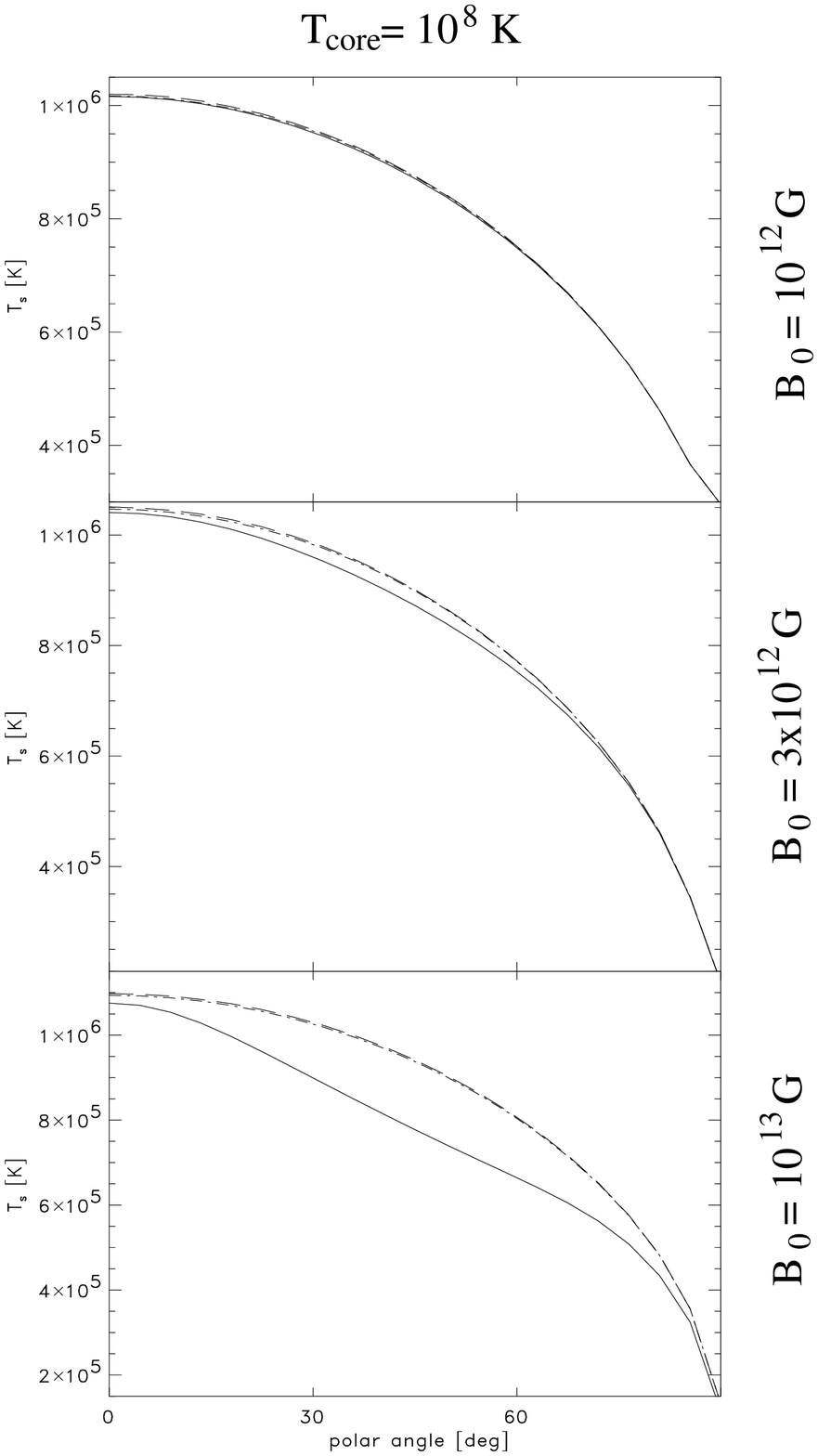}
   \caption{The surface temperature $T_s$ as a function of the polar 
            angle $\theta$ and for
            $T_\mathrm{core} = 10^6$K (left panels),
	    $T_\mathrm{core} = 10^7$K (mid panels), or
	    $T_\mathrm{core} = 10^8$K (right panels) 
            The dashed lines show the surface temperature distribution 
            when an isothermal crust is assumed.
            The full lines represent the surface temperatures when
            the crust temperature distributions take into account
            the anisotropy of heat transport induced by a crustal
            magnetic field (the temperature at the crust-core 
	    interface being fixed at the $T_\mathrm{core}$). 
            Almost indistinguishable from the isothermal crust model is
	    the $T_s$-distribution for a star-centered core field. 
	    It is shown by dot-dashed
	    lines for $T_\mathrm{core} = 10^8$K; for lower $T_\mathrm{core}$ 
	    the differences are even smaller.
            The assumed polar surface field strengths $B_0$ are
            $10^{12}$G (upper panels), $3\cdot 10^{12}$G (mid panels) 
	    and $10^{13}$G (lower panels).
	    }
   \label{fig:surtemp_iso_crust_dip}
\end{figure*}

The magnetic field permeating the envelope induces a non-uniform
surface temperature distribution, mostly due to quantizing effects
of the field at low densities, even in the case of a uniform
crustal temperature (Schaaf \cite{S90a,S90b}; Page \cite{P95}).
The non-isothermality of the crust produced by a crustal magnetic
field will result in an even more pronounced non-uniformity of the
surface temperature.
These effects are shown quantitatively in Fig.~\ref{fig:surtemp_iso_crust_dip}
where the two cases of core and crustal fields are compared.

The different field structures do not only affect the relation between
polar and equatorial surface temperature but also the setup and the 
extension of the warm polar regions.
In Fig.~\ref{fig:surtemp_iso_crust_dip} it is seen that the crustal field can
cause a much smaller warm polar region than a star-centered core field of the
same polar surface strength would do. 
While for the core field geometry with $B_0=10^{13}$G and 
$T_\mathrm{core}=10^8$K the surface temperature at a polar angle of $31.5^0$
is reduced only to $0.93$ of its polar value, for the crustal field
configuration the corresponding value is $0.82$. 
With decreasing $T_\mathrm{core}$ this difference becomes larger: 
thus for $T_\mathrm{core}=10^6$K the corresponding values for the core and 
the crustal field are $0.93$ and $0.77$, respectively. 
The setup of a clearly distinct warm polar region becomes more and
more pronounced with an increasing magnetization parameter (i.e. increasing
field strength and/or decreasing crustal temperature and/or decreasing impurity
concentration) for a crustal field configuration, while the shape of the surface
temperature distribution is almost unaffected by the magnetization parameter in
case of a core magnetic field.

\subsection{Luminosity}

Having a non-uniform surface temperature distribution, $T_s=T_s(\theta)$,
the effective temperature $T_\mathrm{eff}$ can be calculated, from its
definition, as
\be
4\pi R^2\sigma_{\scriptscriptstyle SB}T_\mathrm{eff}^4 \equiv L =  
      \int \sigma_{\scriptscriptstyle SB} T_s^4(\theta) \mathrm{d}\Sigma,
\ee
where $R$ is the circumferential neutron star radius and 
$\mathrm{d}\Sigma \equiv \sin\theta \mathrm{d}\phi \mathrm{d}\theta $ 
is the surface area element, so that finally the  luminosity is given by
\be
L = 2\pi \sigma_{\scriptscriptstyle SB} R^2 \int^\pi_0 T_s^4(\theta) 
       \sin{\theta}\mathrm{d}\theta.
\ee
The photon luminosities for the neutron star models under consideration 
with different magnetic field structures and strengths are listed
in Table~\ref{ta:lum_8} for a model with $T_\mathrm{core}=10^8$K, 
Table~\ref{ta:lum_7} for $T_\mathrm{core}=10^7$K, and 
Table~\ref{ta:lum_6} for $T_\mathrm{core}=10^6$K.
  
The almost identical luminosities obtained for the isothermal crust
and a crust penetrated by a star-centered core field (Table~\ref{ta:lum_8})
reflect the little effect even strong fields of that structure have 
onto the crustal temperature distribution.
While for  a star-centered magnetic field (both with isothermal and
non-isothermal crust) the photon luminosity increases with increasing
field strength, in neutron stars possessing a crustal magnetic 
field above a certain strength ($\approx 10^{12}G$) the luminosity is 
reduced since then over the major part of the surface the heat 
insulating effect of such a field configuration dominates; 
its $\theta$-component causes strong meridional heat fluxes toward the 
polar region whose area, however, becomes smaller with increasing field 
strength. 
This effect impedes the radial heat transport strongly, finally less 
heat can be irradiated away from the surface and the photon cooling process
will be decelerated significantly in comparison to a non-magnetized 
neutron star (here $B_0 \le 10^{12}$G) or even to a strongly
magnetized neutron star with a star-centered core field. 

\begin{table}[t] 
\begin{center} 
\begin{tabular}{cccc}
   B[G]   &     L[erg/s]        &     L[erg/s]        &   L[erg/s]       \\
          &    Isothermal       &  Star-centered      &   Crustal        \\
          &       crust         &      field          &   field          \\
\hline
10$^{12}  $ & $1.66\cdot 10^{32}$ &  $1.64\cdot 10^{32}$ & $1.63\cdot 10^{32}$\\
10$^{12.5}$ & $1.85\cdot 10^{32}$ &  $1.83\cdot 10^{32}$ & $1.68\cdot 10^{32}$\\
10$^{13}  $ & $2.20\cdot 10^{32}$ &  $2.17\cdot 10^{32}$ & $1.25\cdot 10^{32}$\\
\hline
\end{tabular}
\end{center}
\caption[Luminosities]
         {Photon luminosities for a neutron star with $T_\mathrm{core}=10^8$K. The
	 differences between isothermal crust and star-centered field are
	 negligible but significant between them and crustal fields larger than
	 $10^{13}$G.
         \label{ta:lum_8}}
\end{table}

\begin{table}[t] 
\begin{center} 
\begin{tabular}{ccc}
B[G]      &     L[erg/s]        &   L[erg/s]       \\
          &    Isothermal       &   Crustal        \\
          &       crust         &   field          \\
\hline
10$^{12}  $ & $1.42\cdot 10^{30}$ &  $1.31\cdot 10^{30}$\\
10$^{12.5}$ & $1.73\cdot 10^{30}$ &  $1.19\cdot 10^{30}$\\
10$^{13}  $ & $5.01\cdot 10^{30}$ &  $1.00\cdot 10^{30}$\\
\hline
\end{tabular}
\end{center}
\caption[Luminosities]
         {Photon luminosity for a neutron star with $T_\mathrm{core}= 10^7$K.
         \label{ta:lum_7}}
\end{table}

\begin{table}[t] 
\begin{center} 
\begin{tabular}{ccc}
B[G]      &     L[erg/s]        &   L[erg/s]       \\
          &    Isothermal       &   Crustal        \\
          &       crust         &   field          \\
\hline
10$^{12}  $ & $3.36\cdot 10^{28}$ &  $2.82\cdot 10^{28}$\\
10$^{12.5}$ & $4.54\cdot 10^{28}$ &  $2.71\cdot 10^{28}$\\
10$^{13}  $ & $6.92\cdot 10^{28}$ &  $2.59\cdot 10^{28}$\\
\hline
\end{tabular}
\end{center}
\caption[Luminosities]
         {Photon luminosity for a neutron star with $T_\mathrm{core}= 10^6$K.
         \label{ta:lum_6}}
\end{table}

\section{Discussion and conclusions}
\label{sec:discon}

The insulating effect of a quantizing magnetic field component
perpendicular to the radial heat flux is well known and its
role in the low density layers of the envelope of a neutron star
has been extensively studied (e.g., Hernquist \cite{H85}; 
Schaaf \cite{S90a}; Heyl \& Hernquist \cite{HH98}; YP01). 
Here, importance of the classical (Larmor) effect for the heat transport
through the whole crust is demonstrated, provided the neutron star possesses 
a sufficiently 
strong magnetic field maintained by electric currents circulating in
the crust but which does not penetrate the core of the star.
Besides its insulating effect the tangential crustal magnetic field
creates a meridional heat flux from the equatorial regions towards the 
high latitude ones. 
It transports the heat, which is dammed in the equatorial region,
towards the poles where it can be much more easily irradiated away. 
Eventually this leads to an equatorial belt which is much cooler than 
the poles. 
These are a direct consequence of the confinement of the magnetic field
lines to the crust which result in the presence of a large meridional
field component in most of the crust.

Recently, Svidzinsky (\cite{S03}) argued that the accumulation of 
magnetic field lines along the proton superconductor at the crust-core
boundary, due to the Meissner-Ochsenfeld effect, produces an insulating
barrier preventing heat to flow between the crust and the core.
Our results are in the same line of thought but they do not
confirm his claims of insulation of the crust from the core.
The field geometries, i.e., the Stoke's stream functions 
(Eqs.~\ref{equ:Stokes-r},~\ref{equ:Stokes-theta}), 
we used in our calculations come from
field evolution models (Page et al \cite{PGZ00}) of crustal fields
in which the migration, and accumulation, of the currents and the field 
lines toward the crust-core superconducting boundary was modeled in detail 
and, as illustrated in Fig.~\ref{fig:heat_field}, they allow heat diffusion
through the crust-core boundary.
Nevertheless, other field evolution scenarios, e.g. expulsion of the 
magnetic flux from the core by a proton type I superconductor
(Link \cite{L03}; Buckley et al \cite{BMZ04}) may produce a much stronger
piling up of field lines tangentially to the crust-core boundary
and result in more efficient thermal insulation.

In case the magnetic field is allowed to penetrate the core of the star, and 
assuming a star-centered dipolar geometry in the crust, we have shown
that the stationary thermal state of the crust is very close to 
isothermality thus confirming, for this field geometry, the assumptions
of the models of surface temperature distribution of magnetized neutron
stars (e.g., Greenstein \& Hartke \cite{GH83} ; Page \cite{P95}; 
Page \& Sarmiento \cite{PS96}; Shibanov \& Yakovlev \cite{SY96})
which considered only the insulating effect of a quantizing field
in the envelope and assumed the rest of the crust to be isothermal.

The obvious difference between the isothermal crust and our crustal 
field results is demonstrated in Fig.~\ref{fig:b3e12_cru_dens} and
\ref{fig:b3e12_10_8_dip_dens}.
For a crustal field these differences cannot be neglected for 
highly magnetized neutron stars. 

Since the warm polar region in case of a strong crustal field is much
smaller than for star-centered and/or weak magnetic fields 
(see Fig.~\ref{fig:surtemp_iso_crust_dip}) this may open a new way
to distinguish between crustal and core magnetic fields:
{\em A strong crustal magnetic field implies a smaller effective 
area for thermally emitting cooling neutron stars}.
This has consequences which can potentially be observed in X-ray
in cases where fits of the thermal component of the spectrum of a cooling
neutron star result in effective emitting radii $R_\mathrm{eff}$ which are
significantly smaller than the expected radius of a neutron star.
For example, the ``Three Musqueteers''
(PSR 0656+14, PSR 1055-52 and Geminga: Tr\"umper \& Becker \cite{TB98})
as well as RX J185635-3754 (Pons et al \cite{Petal02}) all 
have $R_\mathrm{eff} \sim 5-7$ km when their thermal spectra are fitted
with blackbody spectra.
If these radii would coincide with the radius of the neutron star, 
the equation of state describing the state of the core matter would 
have to be extremely soft.
A relatively small warm polar region, created by a strong 
crustal field and emitting almost all the thermal radiation would be 
a reasonable explanation for such small  $R_\mathrm{eff}$. 

The differences in the photon luminosities for a star-centered or a crustal 
field will also affect the long term cooling of neutron stars.
Future cooling calculations as well as the comparison of X-ray 
spectra and pulse profiles with model calculations assuming different 
field structures will open a way to discriminate the two basic 
scenarios: crustal or core magnetic field. 
Finally we mention the consequences the non-isothermality of the 
crust may have for the crustal field itself. 
Since the electric conductivity in the hot polar region is much 
smaller than in the equatorial layer, the field decay will be
affected too and may cause differences in the field structure. 
This ``back reaction'' of the field onto its own decay via a field 
driven non-spherical symmetric crustal temperature distribution is 
subject of future investigations.

\begin{acknowledgements}
U.G. is grateful to G. Ruediger, whose engagement enabled the 
realization  of the DFG-project 
"The interaction of thermal and magnetic effects in neutron stars" 
(RU 488/18-1).
Part of this is supported by a binational grant from DGF-Conacyt
\#444MEX113/4/0-2.
D.P.'s work is partially supported by grants from
UNAM-DGAPA (\#IN112502) and Conacyt (\#36632-E).
\end{acknowledgements}



\end{document}